\documentclass[runningheads]{llncs}
\usepackage[T1]{fontenc}
\usepackage{amsmath}
\usepackage{graphicx}
\usepackage{listings}
\usepackage[dvipsnames]{xcolor}
\usepackage{float}
\usepackage{xurl} 
\usepackage{hyperref}
\usepackage{amssymb}
\usepackage{cleveref}
\usepackage{booktabs}
\usepackage{xcolor}
\definecolor{highlight}{RGB}{255,255,150}  
\definecolor{redhighlight}{RGB}{255,180,180}  
\usepackage{multirow} 
\begin{document}
\pagestyle{headings}
\title{Feature Toggle Dynamics in Large-Scale Systems: Prevalence, Growth, Lifespan, and Benchmarking}
\author{Xhevahire Tërnava\orcidID{0000-0002-4317-2764}}
\institute{LTCI, Télécom Paris, Institut Polytechnique de Paris, Palaiseau, France \\
\email{xhevahire.ternava@telecom-paris.fr}}

\authorrunning{Xh. Tërnava}

\titlerunning{Feature Toggle Dynamics in Large-Scale Systems}
%
%
%
%
\maketitle              
\begin{abstract}
Feature toggles enable gradual rollouts and experimentation in software systems, yet often persist beyond their intended lifecycle, accumulating as technical debt. Prior research has examined feature toggle interactions and complexity, but no longitudinal study has quantified how toggles evolve over time across different organizational contexts. We analyse over 4,000 toggle events in Kubernetes (10 MLoC, 8.5 years) and GitLab (5 MLoC, 5 years). We find that feature toggle removals lags behind additions in both systems (by roughly 35\% and 13\%, respectively), leading to growing toggle inventories. Their lifespan patterns also differ notably, with Kubernetes toggles lasting a median of 734 days versus 185 in GitLab. Then, some feature toggles (1.33\% and 0.73\%, respectively) exceed all previously observed removal durations, becoming de facto permanent. Building on these findings, we propose a benchmarking framework with five key metrics and their empirically derived threshold zones, enabling practitioners to assess and compare toggle management practices across projects. All scripts and data are publicly available.
\end{abstract}
\section{Introduction}
\label{introduction}
Feature toggles have become an essential part of modern software development~\cite{github_featureflags_2021}. They are Boolean switches~\cite{hodgson2017feature} that allow teams to deploy unfinished features, run experiments on live traffic, and roll back problematic changes, all without touching the deployment pipeline. In theory, toggles are meant to be temporary~\cite{meinicke2020exploring}. Once a toggled feature is completed, the toggle should be removed and the code cleaned up. In practice, things rarely work out that way.

The problem is that feature toggles are easy to add and easy to forget~\cite{reddit_implementation_details_2024}. A common scenario unfolds when a developer adds a toggle to guard a new feature in the system, during the time the feature stabilizes, and the toggle quietly remains in the codebase. Months later, nobody remembers why it exists or who owns it. Thus, the toggle becomes permanent not by design, but by neglect. If we multiply this across a large codebase, we may end up with dozens, sometimes hundreds~\cite{ramanathan2020piranha}, of toggles accumulating like sediment in the codebase. Each one adds conditional logic and makes the system harder to test and understand.

Despite the widespread use of feature toggles~\cite{flickr2009,meinicke2020capture,prutchi2021adoption,rahman2023feature,schroder2022discovering}, we know surprisingly little about their actual lifecycle in production systems. Existing studies and gray literature address feature toggle interactions~\cite{schroder2022discovering,ternava2022interaction}, added complexity~\cite{rahman2024exploring}, usage patterns~\cite{rahman2023feature}, addition/removal practices~\cite{Hezaveh2021,neely2013continuous,rahman2016feature}, underlying rationales~\cite{hodgson2017feature,meinicke2020exploring}, management~\cite{unleash2024,meinicke2020capture,roy2021}, and standardization~\cite{openfeature2025}. Yet key questions remain unexplored: 
\textit{How many feature toggles exist in modern codebases?} \textit{How fast are they added compared to how fast they are removed?} \textit{Do active feature toggles accumulate over time, or do teams keep up with cleanup?} And importantly, \textit{how long do individual feature toggles survive before someone finally deletes them?} These questions matter because the answers determine whether feature toggle management is sustainable at scale or whether teams are slowly drowning in technical debt~\cite{neely2013continuous,ramanathan2020piranha} they cannot see.

This paper provides empirical answers to such questions. We conducted a longitudinal study of feature toggles in two large-scale open source systems, namely Kubernetes, a container orchestration platform with roughly 10 million lines of Go, and GitLab, a DevOps application with about 5 million lines of Ruby. By mining over 8.5 and 5 years, respectively, of their version control history, we tracked when toggles were added, when they were removed, and which ones are still alive despite exceeding all historical precedents. We find that feature toggle lifecycles vary greatly across projects. Basically, what counts as ``overdue'' in one codebase may be perfectly normal in another. We also find that cleanup of toggles rarely keeps pace with their addition, leading to a gradual accumulation of active toggles over time (about 35\% and 13\%, respectively).

Moreover, building on these findings, we propose the first benchmarking framework for feature toggle management. The framework introduces five metrics that quantify toggle events, 
accumulation rate, cleanup effectiveness, codebase density, and relative lifespan, enabling systematic cross-project comparison. We then derive their threshold zones empirically using Kubernetes and GitLab as baselines. The resulting framework offers practitioners concrete means to assess toggle health and benchmark their practices against real-world baselines.

In summary, this paper makes the following contributions:
\begin{itemize}
    \item A longitudinal study of feature toggle dynamics in two large-scale systems, covering over 4{,}000 toggle events across more than 8 years of development. 
    \item Empirical evidence that toggle lifecycles are project-dependent, with \textit{e.g.,} median lifespans differing by nearly four times between software systems.
    \item A benchmarking framework with five complementary metrics and threshold zones, enabling practitioners to assess toggle health across projects. An interactive dashboard is also available at \href{https://ternava.github.io/detog/}{https://ternava.github.io/detog/}.
    \item A replication package containing all scripts and data is publicly available at \href{https://doi.org/10.5281/zenodo.18773811}{https://doi.org/10.5281/zenodo.18773811} to enable reproducibility.
\end{itemize}

The rest of this paper is structured as follows. \Cref{backgroundandmotivation} provides background on feature toggles and the need for empirical lifecycle studies, \Cref{experimentalapproach} describes the study design, and \Cref{results} presents the findings. \Cref{benchmarking} introduces the benchmarking framework, followed by threats to validity in \Cref{threatstovalidity}, related work in \Cref{relatedwork}, and conclusion with future directions in \Cref{conclusion}.

\begin{lstlisting}[
  language=Go,
  frame=single,
  numbers=left,
  numberstyle=\tiny\color{gray},
  basicstyle=\ttfamily\scriptsize,
  keywordstyle=\color{blue}\bfseries,
  commentstyle=\color{ForestGreen}\itshape,
  stringstyle=\color{orange},
  breaklines=true,
  showstringspaces=false,
  escapeinside={(*@}{@*)},
  caption={Feature gate definition in Kubernetes, \href{https://github.com/kubernetes/kubernetes/blob/61c629cc57649bf0ce2378903027c83939d84109/pkg/features/kube_features.go\#L62}{/pkg/features/kube\_features.go}},
  label=listing1,
]
const (
// owner: @bswartz
// Enables usage of any object for volume data source in PVCs
(*@\colorbox{highlight}{AnyVolumeDataSource featuregate.Feature="AnyVolumeDataSource"}@*)
//... the rest of feature gate definitions
)
var defaultVersionedKubernetesFeatureGates = map[featuregate.Feature]featuregate.VersionedSpecs {
    (*@\colorbox{highlight}{AnyVolumeDataSource}@*): { // ...
    {Version: version.MustParse("1.33"), Default: true, PreRelease: featuregate.GA, LockToDefault: true}, (*@\colorbox{redhighlight}{// GA in 1.33 -> remove in 1.36}@*)
    }, // ...
}
\end{lstlisting}
\begin{lstlisting}[
  language=Go,
  frame=single,
  numbers=left,
  numberstyle=\tiny\color{gray},
  basicstyle=\ttfamily\scriptsize,
  keywordstyle=\color{blue}\bfseries,
  commentstyle=\color{ForestGreen}\itshape,
  stringstyle=\color{orange},
  breaklines=true,
  showstringspaces=false,
  escapeinside={(*@}{@*)},
  caption={Feature gate usage, \href{https://github.com/kubernetes/kubernetes/blob/61c629cc57649bf0ce2378903027c83939d84109/pkg/api/persistentvolumeclaim/util.go\#L48}{/pkg/api/persistentvolumeclaim/util.go}},
  label=listing2,
]
// Drop the contents of the dataSourceRef field if the AnyVolumeDataSource feature gate is disabled.
(*@\colorbox{highlight}{if !utilfeature.DefaultFeatureGate.Enabled(features.AnyVolumeDataSource)}@*) {
    if !dataSourceRefInUse(oldPVCSpec) {
        pvcSpec.DataSourceRef = nil
    }
}
\end{lstlisting}

\section{Background and Motivation}
\label{backgroundandmotivation}
Feature toggles are simple Boolean switches embedded in codebase that enable or disable functionality at runtime, allowing teams to deploy incomplete features, conduct A/B experiments, and control rollouts without redeploying~\cite{flickr2009,hodgson2017feature}. They have become particularly valuable in modern Continuous Integration and Continuous Delivery (CI/CD) pipelines~\cite{humble2010continuous}, where they decouple deployment from release and enable safer, more frequent deliveries.
\Cref{listing1} illustrates how feature toggles (also known as \textit{feature gates} in Kubernetes) are defined within a codebase, using \texttt{AnyVolumeDataSource} as an example (line 4), while \Cref{listing2} shows their conditional usage within that codebase.
Notably, Kubernetes embeds explicit removal timelines in feature gate definitions (comment in line 9 in \Cref{listing1}), indicating when a gate should be removed after the guarded feature is stabilized. 
This practice supports proactive lifecycle management.
While \textit{temporary} spans a wide range of durations, and some toggles can be explicitly designated as permanent (\textit{e.g.,} as in the case of Chromium: \href{https://chromium.googlesource.com/chromium/src/+/refs/heads/main/chrome/browser/flag-never-expire-list.json}{\texttt{flag-never-expire-list}}), the toggles in Kubernetes and GitLab are explicitly designed for removal. Kubernetes gates follow a staged deprecation process, and GitLab documentation prescribes cleanup after feature stabilization.
Hence, originally designed as temporary scaffolding for gradual delivery, these feature toggles are supposed to be removed (\textit{i.e.,} lines 2-10 in~\Cref{listing1} and 1-2, 6 in~\Cref{listing2}) once a toggled feature is completed (\textit{cf.} lines 3-5 in~\Cref{listing2}) and the code cleaned up~\cite{meinicke2020exploring}. In practice, however, they often linger. Each forgotten toggle adds conditional logic that increases cyclomatic complexity~\cite{ramanathan2020piranha}, expands the state space requiring testing~\cite{rahman2024exploring}, and makes the codebase harder to understand and reason about. Prior research has examined toggle interactions~\cite{schroder2022discovering,ternava2022interaction} and complexity~\cite{rahman2024exploring}, 
yet, to the best of our knowledge, there remains no systematic empirical evidence on \textit{toggle prevalence in modern codebases, their temporal accumulation patterns, and their actual survival duration in production}.

Without empirical data on these dynamics, teams cannot tell if their toggle inventory is growing out of control, whether a six-month-old toggle is normal or overdue, or whether cleanup keeps pace with addition. To the best of our knowledge, they lack benchmarks to distinguish healthy additions / removals of toggles from silent accumulation, making evidence-based removal policies impossible. In fact, practitioners have long recognized this feature toggles accumulation problem, some have even proposed a ``one-in-one-out'' rule, requiring the removal of an existing toggle before adding a new one~\cite{hodgson2017feature}. 
Others advocate embedding explicit expiration dates directly in toggle definitions (\textit{cf.}\ \Cref{listing1}, line 9) to enable timely and potentially automated cleanup.
Yet, without quantitative data, teams cannot assess for instance when, at what age of their project, to apply such rule. This challenge may also intensify as AI-assisted development accelerates feature delivery~\cite{github2025octoverse02}, shortening iteration cycles and growing toggle volumes~\cite{github2025octoverse01}. 
This gap motivates both empirical investigation and the establishment of quantitative benchmarks that help teams assess their toggle health, compare practices across organizational contexts, and determine when intervention is needed.

\section{Study Design}
\label{experimentalapproach}
%

\subsection{Research Questions}
\label{researchquestions}
To empirically characterise feature toggle lifecycle dynamics, we address three research questions using data from Kubernetes~\footnote{Kubernetes codebase, \url{https://github.com/kubernetes/kubernetes}.} and GitLab~\footnote{GitLab codebase, \url{https://gitlab.com/gitlab-org/gitlab}.}.

\begin{description}
    \item[$RQ_{1}$:]\label{RQ1} \textbf{How do the counts of feature toggles added and removed evolve over the lifecycle of software projects?}
        We track addition and removal counts over time to assess whether cleanup keeps pace with additions and to identify bulk events (refactoring, merging) that produce significant spikes, revealing whether toggles accumulate silently or are actively managed.
    \item[$RQ_{2}$:]\label{RQ2} \textbf{How does the number of active feature toggles change over the lifecycle of software projects?}
        We compute the cumulative sum of additions minus removals over time, normalised by lines of code, to assess whether active toggle inventories grow, stabilise, or decline across projects.
    \item[$RQ_3$:]\label{RQ3} \textbf{How do the lifespans of feature toggles differ between software projects?}
        We measure the time from toggle addition to removal using Kaplan-Meier survival analysis and classify toggles by project-specific lifespan quartiles, identifying whether lifecycle patterns are universal or project-dependent and flagging de facto permanent toggles within a project.
\end{description}

\subsection{Analysed Projects}
\label{analysedprojects}
To address our research questions, we analysed 
the feature toggles in two real-world software systems, namely Kubernetes and GitLab. These projects are suitable subjects because they are large, actively maintained, widely adopted, and differ significantly in architecture. Kubernetes is a Go-based container orchestration platform and GitLab is a Ruby on Rails DevOps application, with approximately 10 million (9.98M LoC) and 5 million (4.86M LoC) lines of code, respectively, in the last analysed version of them~\footnote{Measured with \texttt{cloc}, \url{https://github.com/AlDanial/cloc}.}. 
Their release cycles also differ. Kubernetes releases approximately quarterly while GitLab releases monthly.

The way feature toggles are defined and used in these systems also differs. Kubernetes implements toggles as \textit{feature gates}, defined in the \href{https://github.com/kubernetes/kubernetes/blob/master/pkg/features/kube\_features.go}{pkg/features/kube\_features.go} file, whereas GitLab implements them as \textit{feature flags}, defined in dedicated YAML files under the \href{https://gitlab.com/gitlab-org/gitlab/-/tree/master/config/feature\_flags?ref_type=heads}{\texttt{config/feature\_flags/}} directory. 
Throughout this paper, we usually use the terms \textit{gates} and \textit{flags} when referring specifically to Kubernetes and GitLab feature toggles, respectively.
Moreover, in this study, we analysed the commit histories of Kubernetes from June 06, 2014 (\href{https://github.com/kubernetes/kubernetes/commit/2c4b3a562ce34cddc3f8218a2c4d11c7310e6d56}{2c4b3a5}) to August 19, 2025 (\href{https://github.com/kubernetes/kubernetes/commit/4e8b192b66cc2a6952b8f1a5067e563c4019c276}{4e8b192}), covering about 11 years and 2 months, and of GitLab from October 09, 2011 (\href{https://gitlab.com/gitlab-org/gitlab/-/commit/93efff945215a4407afcaf0cba15ac601b56df0d}{93efff94}) to August 22, 2025 (\href{https://gitlab.com/gitlab-org/frontend/gitlab-ui-integrations/-/commit/8127b1c5bc226ca6d422a3dadceaf59189ac136b}{8127b1c5}), covering about 13 years and 10 months.

Their historical trajectories also differ. Kubernetes was started on June 06, 2014, but on January 20, 2017, it introduced for the first time 5 feature gates in the main branch (which we analysed), about 2.5 years after its start. However, the GitLab repository was created on October 09, 2011, but its first feature flag appeared in the main branch only on July 08, 2020, almost 9 years later. This means we analysed in this study approximately 8 years and 7 months of feature gate history in Kubernetes and 5 years and 1 month of feature flag history in GitLab, up to August 2025. Currently, Kubernetes has 155 active feature gates, and GitLab has 403 active feature flags, all of which are Boolean.

\subsection{Mining Feature Toggles}
\label{miniming}
Typically, feature toggles are defined in a central location within a software system, either at the statement level or the file level. For instance, in Kubernetes, feature gates are declared as constants within specific files, such as the \texttt{AnyVolumeDataSource} gate in the \href{https://github.com/kubernetes/kubernetes/blob/master/pkg/features/kube\_features.go}{pkg/features/kube\_features.go} file shown in~\Cref{listing1} (line 4). As for GitLab, feature flags are defined in their own YAML files organized within dedicated directories, such as for example the \href{https://gitlab.com/gitlab-org/gitlab/-/blob/master/config/feature_flags/ops/access_rest_chat.yml?ref_type=heads}{\texttt{access\_rest\_chat}} flag in the \href{https://gitlab.com/gitlab-org/gitlab/-/tree/master/config/feature_flags?ref_type=heads}{\texttt{config/feature\_flags/ops/}} directory.

To analyse the evolution of feature toggles, specifically the addition of new toggles or removal of existing ones, we extracted and processed the version control history of the corresponding files and directories. For Kubernetes, we examined the commit history of the \href{https://github.com/kubernetes/kubernetes/blob/master/pkg/features/kube\_features.go}{pkg/features/kube\_features.go} file, where most of the gates are defined. For GitLab, we analysed the commit history of all eight subdirectories under the \href{https://gitlab.com/gitlab-org/gitlab/-/tree/master/config/feature_flags?ref_type=heads}{config/feature\_flags/} directory.
For each file and directory, we parsed the commit logs and identified toggle additions and removals through pattern matching on the unified diff output. Lines prefixed with \texttt{+} indicated additions, and lines prefixed with \texttt{-} indicated removals. For Kubernetes, we matched feature gate declarations of the form \texttt{MyFeature featuregate.Feature = "MyFeature"}, whereas for GitLab, we tracked the creation and deletion of \texttt{*.yml} files.
Specifically, we recorded the name of the changed gate or flag, whether the change represented an addition or a removal, the commit SHA, and the associated timestamp. These data served as inputs for our subsequent analysis, including determining the active status and lifespan of individual feature toggles. 
To validate the accuracy of this automated extraction, we cross-checked the results by manually inspecting the commit logs of ten randomly selected toggles from each system and comparing the obtained active toggles against the listed toggles in the project documentation and command-line outputs (for this, we used \texttt{kube-apiserver} -{}-\texttt{help} for Kubernetes, and \texttt{find config/feature\_flags \mbox{-}name '*.yml' | wc -l} for GitLab).

Moreover, to ensure full reproducibility, all build scripts, processed data, and results are publicly available at~\href{https://doi.org/10.5281/zenodo.18773811}{https://doi.org/10.5281/zenodo.18773811}.

\section{Results}
\label{results}
This section presents the results obtained regarding the research questions.

\subsection{Added and Removed Feature Toggles}
\label{addedremovedfeaturetoggles}
\begin{figure}[t]
    \centering
    \includegraphics[width=1\linewidth]{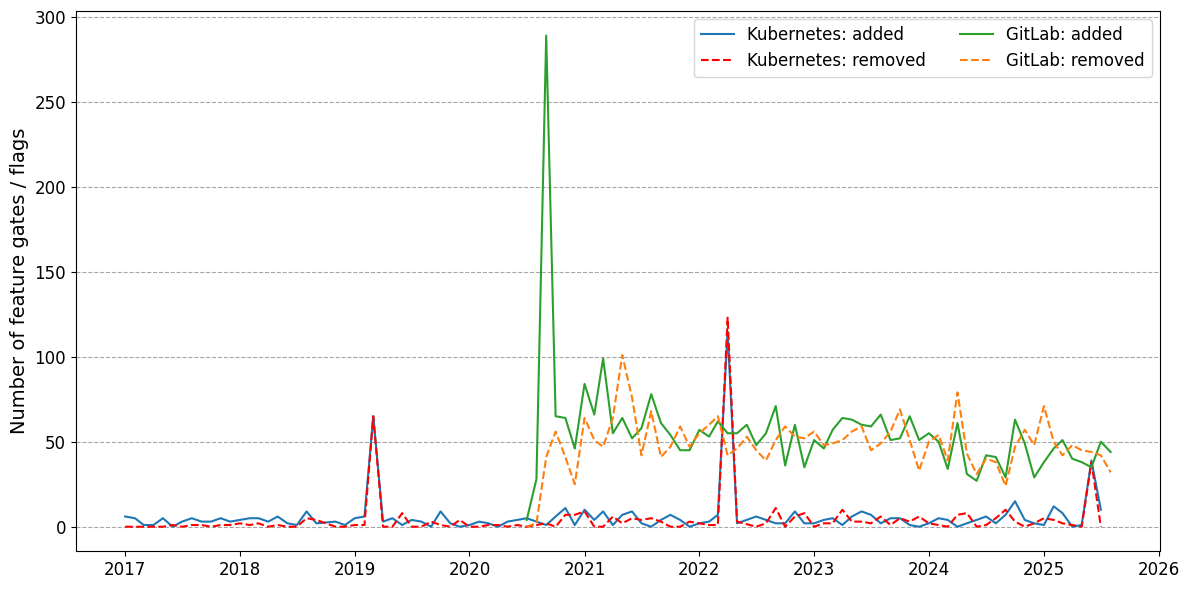}
    \caption{Added vs. removed feature gates in Kubernetes and flags in GitLab}
    \label{fig:gitaddedvsremoved}
\end{figure}
As a first step, we examined the number of feature toggles added and removed over time in both subject systems. \Cref{fig:gitaddedvsremoved} presents the monthly totals of changes for Kubernetes and GitLab. The analysis revealed that 603 feature gates were added over 8 years and 7 months in Kubernetes, of which 448 were subsequently removed within the same period. Similarly, over 5 years and 1 month in GitLab, 3,442 feature flags were added, and 3,038 were later removed.

When comparing the line charts of added and removed toggles in \Cref{fig:gitaddedvsremoved}, it can be observed that, for both systems, the numbers of added and removed toggles generally evolve in parallel. 
This suggests that most toggles added are eventually removed, maintaining a relatively consistent balance between additions and removals over time. There are, however, a few notable exceptions. Specifically, there are three major bulk additions and removals of feature gates in Kubernetes, and one major bulk addition of feature flags in GitLab.
The three bulk changes in Kubernetes were triggered by refactoring. The first occurred in early 2019, when the definition syntax of feature gates was updated. Namely, the old format \texttt{MyFeature utilfeature.Feature = "MyFeature"} was replaced by \texttt{MyFeature featuregate.Feature = "MyFeature"}, resulting in the refactoring of 65 gates. The second took place in April 2022, when all feature gates were reordered alphabetically to reduce merge conflicts. On the same day, several mature gates were removed, leading to 112 additions and 123 removals. The most recent bulk change, in June 2025, also involved alphabetizing, during which 34 gates that were not properly sorted were adjusted.
In GitLab, a significant number of feature flags were added toward the end of 2020, when 222 development feature flags were merged into the master branch as part of a migration process, effectively incorporating nearly all development flags into the main codebase.

On average, 5{.}85 feature gates are added per month in Kubernetes and 55.52 feature flags in GitLab, while 4.35 gates and 49.00 flags are removed, respectively. These averages indicate that, slightly  more feature toggles are added than removed (by 34.48\% and 13.31\% each month, respectively). The differences in these averages suggest that the rate of change in amount and frequency varies across projects. A closer investigation showed that they are influenced by factors such as refactoring, release phases, and feature toggle management practices.
\\

\noindent\setlength{\fboxsep}{6pt} 
\fbox{%
\parbox{0.95\linewidth}{%
\hyperref[RQ1]{$RQ_{1}$} \textbf{insights:}
Feature toggles are regularly added and removed in software projects. However, the frequency varies significantly across projects 
(GitLab has 10 times more toggle events per month than Kubernetes).
Despite these differences, the roughly parallel evolution of additions and removals within each project indicates that teams actively manage feature toggle lifecycles, balancing experimentation with maintainability even as a small net accumulation of toggles persists (35\% surplus in Kubernetes, 13\% in GitLab).
  }%
}

\subsection{Active Feature Toggles}
\label{activefeaturetoggles}
\begin{figure}[t]
    \centering
    \includegraphics[width=1\linewidth]{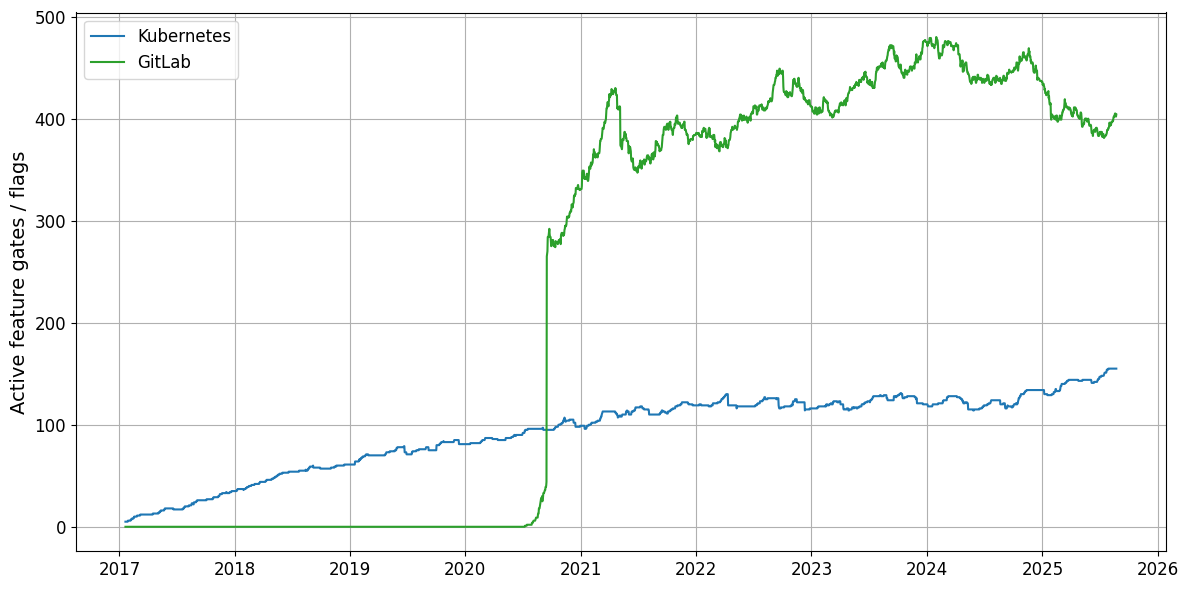}
    \caption{Daily active toggles in Kubernetes and GitLab, cumulative values}
    \label{fig:cubeactiveftoertime}
\end{figure}
To address our second research question, we analysed the number of feature toggles that were active in the codebases on each day. 
We did this by calculating the cumulative sum of added minus removed toggles in both subject systems, which allowed us to track whether their numbers increased, stabilized, or decreased over the project’s history. We first noted that in the last analysed commits, 155 feature gates in Kubernetes and 403 feature flags in GitLab are active.

\Cref{fig:cubeactiveftoertime} shows the evolution of active feature toggles in both Kubernetes and GitLab. We observed that Kubernetes started with 5 active gates on January 20, 2017. Over the course of that year, the number grew almost linearly, reaching a maximum of 35 active gates on December 22, 2017. This almost linear growth continued for about two years. Then, beginning in March 2019, the number of active gates in Kubernetes rose steadily from 70 to 155 by July 29, 2025, reflecting a net growth rate of approximately 0.036 gates per day (about 13.3 per year) over 2,339 days. We observed that this growth pattern was shaped by multiple additions between release cycles and subsequent removals prior to releases, producing visible stepwise changes in the cumulative totals. 
In contrast, ~\Cref{fig:cubeactiveftoertime} shows that GitLab's active feature flags increased sharply from 1 on July 8, 2020, to 335 within the same year. Growth persisted until reaching 403 active flags on August 23, 2025, before showing signs of stabilization and a slight recent decline. This translates to an average net growth rate of about 0.215 flags per day (approximately 78.6 per year) across 1,872 days in GitLab.

Despite their differing growth trajectories, both systems maintain considerable operational feature toggle surfaces. On average, Kubernetes has 102.6 active toggles per day, compared to 394.2 in GitLab. When normalized by lines of code (LoC), the disparity is more pronounced. Based on the last analysed commits, Kubernetes has one gate per $\approx$64 kLoC, while GitLab exhibits a much denser toggle surface with about one flag per $\approx$12 kLoC. This five fold higher toggle density in GitLab highlights a significant imbalance and underscores the considerable ongoing effort required to manage long-lived toggles in both codebases.
\\

\noindent\setlength{\fboxsep}{6pt} 
\fbox{%
  \parbox{0.95\linewidth}{%
\hyperref[RQ2]{$RQ_{2}$} \textbf{insights:}
Active feature toggles generally accumulate over a software project's lifecycle, though growth rates vary significantly (from 13 to 79 feature toggles per year in our study). When normalized by codebase size, feature  toggle density can differ fivefold across projects, reflecting different release cycles, organizational practices, and maintenance overhead.
  }%
}

\begin{figure}[t]
    \centering
    \includegraphics[width=1\linewidth]{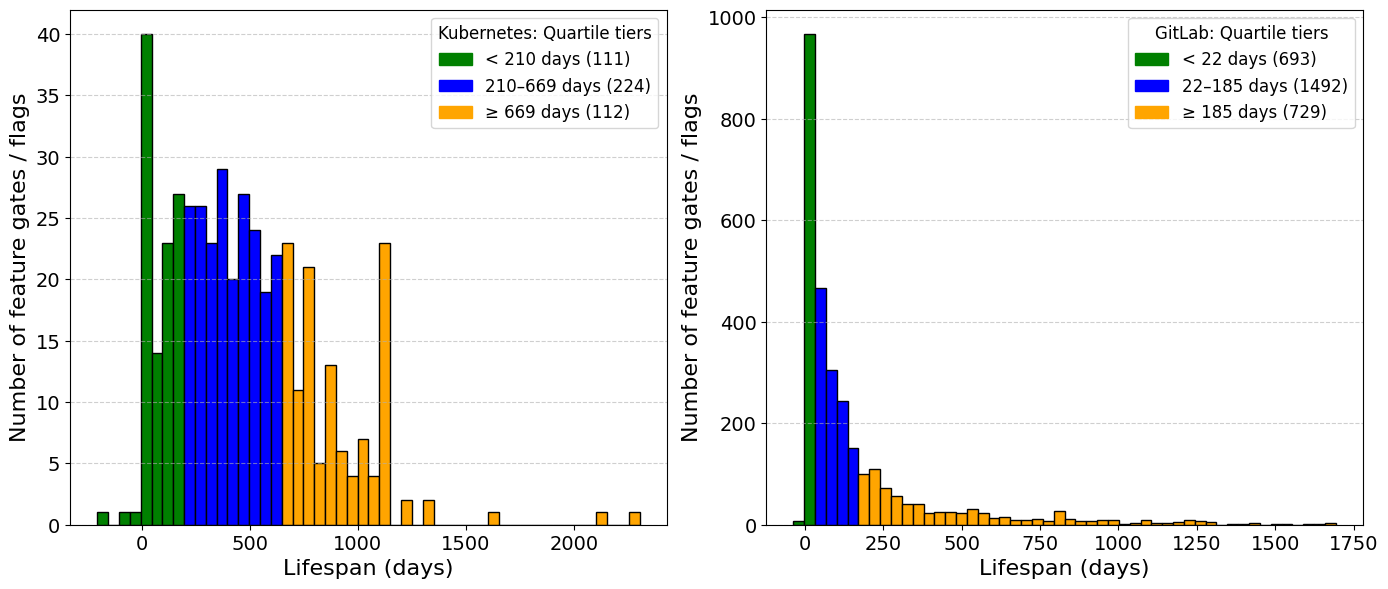}
    \caption{Distribution of feature toggles longevity in Kubernetes and GitLab}
    \label{fig:cubelongevity}
\end{figure}
\begin{figure}[t]
    \centering
    \includegraphics[width=\linewidth]{assets/05-git-kube-KM-survival-curves.png}
    \caption{Kaplan-Meier survival curves showing feature toggle lifespans in Kubernetes and GitLab. Active toggles are marked with vertical ticks, whereas red marks indicate those exceeding the maximum observed lifespan of removed toggles}
    \label{fig:cubeactivelongevity}
\end{figure}

\subsection{The Lifespan of Feature Toggles}
\label{lifetime}
To address our third research question, we analysed the lifespan of feature gates in Kubernetes and feature flags in GitLab across their entire project histories. Specifically, we measured the time elapsed between the addition and removal of each feature toggle. Then, we classified toggles into three categories: \textit{temporary}, \textcolor{Green}{$\blacksquare$} (removed within the first 25\% of the overall lifespan), \textit{intermediate}, \textcolor{blue}{$\blacksquare$} (removed between 25\% and 75\% of the overall lifespan), \textit{long-lived}, \textcolor{orange}{$\blacksquare$} (removed at or beyond 75\% of the overall lifespan), and \textit{permanent}, \textcolor{red}{\rule{1.5pt}{0.8em}} (never removed, or with a lifespan exceeding that of the longest-lived feature gate or flag).

The results for Kubernetes, shown in~\Cref{fig:cubelongevity} (left), indicate that 111 feature gates were \textit{temporary}, 224 had an \textit{intermediate} lifespan, and 112 were \textit{long-lived}. For GitLab (\Cref{fig:cubelongevity}, right), 693 feature flags were \textit{temporary}, 1,492 \textit{intermediate}, and 729 \textit{long-lived}. While the proportions across tiers are similar (approximately 25\%, 50\%, and 25\%) due to quartile-based classification, the key insight lies in the \textit{absolute day thresholds} that define each tier. In Kubernetes, \textit{temporary} means fewer than 210 days, whereas in GitLab it means fewer than 22 days, that is, nearly a tenfold difference. This implies that a toggle considered ``temporary'' in Kubernetes would already qualify as ``long-lived'' in GitLab, underscoring how strongly feature toggle lifecycle expectations differ across projects.

Note that a small number of bars appear before day zero in \Cref{fig:cubelongevity}, corresponding to toggles with negative computed lifespans caused by git history anomalies (\textit{e.g.,} rebasing or cherry-picking across branches). These represent less than 1\% of all removed toggles and do not affect our findings (\textit{cf.}~\Cref{internalthreats}).

The maximum observed toggle lifespan also differs considerably between the two projects, which is 2,305 days (approximately 6.3 years) in Kubernetes versus 1,696 days (approximately 4.6 years) in GitLab. Moreover, the average lifespan is 471.6 days in Kubernetes compared to 165.7 days in GitLab, indicating that toggles in Kubernetes persist nearly three times longer on average.
These differences align with each project's release cadence. Kubernetes gates follow a staged promotion process (Alpha$\to$Beta$\to$GA) spanning roughly quarterly releases, while GitLab flags target removal within monthly cycles. As we show in \Cref{benchmarking}, when lifespan is expressed in release cycles rather than calendar days, both systems fall in the same range of approximately 6 to 7 cycles, suggesting that release cadence accounts for much of the absolute lifespan difference.

To complement the lifespan distribution of removed toggles above, we applied survival analysis to model how long feature toggles remain active before removal. The Kaplan-Meier estimator is well-suited for this purpose because it accounts for \textit{censored} data, that is, toggles still active at the end of our study whose eventual removal date remains unknown~\cite{bland1998survival}. Unlike the histogram analysis in~\Cref{fig:cubelongevity}, which only includes toggles that have been removed, the survival analysis incorporates all toggles, both removed and currently active.
As a result, the survival curves in \Cref{fig:cubeactivelongevity} reveal notable differences in toggle lifecycle management. In Kubernetes, half of all feature gates remain active for at least 734 days (roughly 2 years), whereas in GitLab, the median survival time is only 185 days (about 6 months), nearly four times shorter. This gap is also visible in the shape of the curves, showing that Kubernetes maintains a survival probability above 50\% until around day 700, whereas GitLab's curve drops below this threshold much earlier, suggesting more aggressive or frequent cleanup practices.

Of particular concern are also the toggles represented by red vertical markers in \Cref{fig:cubeactivelongevity}, which denote currently active toggles that have already exceeded the maximum observed lifespan among all historically removed toggles in their respective systems (\textit{i.e.,} 2,305 days in Kubernetes and 1,696 days in GitLab). 
Specifically, we identified eight Kubernetes feature gates and 25 GitLab feature flags that have surpassed these thresholds. 
These outliers represent de facto \textit{permanent} toggles, which have persisted well beyond typical cleanup windows and now risks becoming deeply entangled with the codebase.
For example, Kubernetes's \href{https://github.com/kubernetes/kubernetes/commit/d55f037b7d84e61c81a6b8c20606bb06b6eb20f2}{\texttt{HPAScaleToZero}} gate, which guards Horizontal Pod Autoscaler scale-to-zero behavior, has remained in Alpha for over 2,353 days (since 2019). In GitLab, \href{https://gitlab.com/gitlab-org/gitlab/-/commit/1e15e99b8dbfdae2618244d9c9722e13d752c61b}{\texttt{personal\_snippet\_reference\_filters}}, a development flag from the 2020 migration batch, has been active for over 1,845 days without cleanup.

The existence of such long-lived active toggles raises a critical question: \textit{when does a ``temporary'' experimental feature effectively become permanent?} Our data suggest that once a toggle survives beyond the historical maximum removal threshold, its probability of eventual removal drops considerably. The survival curve analysis thus provides not merely a descriptive mean but a predictive way for those toggles that may exceed the historical norms of their longevity. Moreover, this finding highlights the importance of proactive toggle management practices in software projects, including the use of explicit expiration dates, automated cleanup mechanisms, and regular toggle reviews within development workflows. Such measures can help prevent the accumulation of stale toggles, reduce code complexity, and enhance overall system maintainability.
\\

\noindent\setlength{\fboxsep}{6pt} 
\fbox{%
  \parbox{0.95\linewidth}{%
\hyperref[RQ3]{$RQ_{3}$} \textbf{insights:}
Feature toggle lifespans are highly system dependent, with Kubernetes toggles persisting nearly three times longer on average (472 vs. 166 days) and half of them surviving at least four times longer (734 vs. 185 days). A small but important  subset of active toggles (8 in Kubernetes, 25 in GitLab) has exceeded all historical removal thresholds, effectively becoming permanent and highlighting the need for proactive cleanup mechanisms.
  }%
}

\section{Feature Toggle Benchmarking Framework}
\label{benchmarking}
Our analysis of Kubernetes and GitLab reveals that toggle management practices vary significantly across projects, with differences in churn rates (10 times higher in GitLab), toggle density (5 times higher per kLoC), and median lifespans (4 times longer in Kubernetes). These findings raise a natural question: \textit{how can practitioners determine whether their own toggle inventory is healthy or problematic?} For example, \textit{is 20 active toggles too many for a 200 kLoC project?} or \textit{is adding 10 toggles per month sustainable?} To address these questions, we propose five \textit{benchmark metrics} for cross-project comparison of toggle management practices, together with an \textit{interpretation framework} consisting of empirically derived threshold zones using Kubernetes and GitLab as baselines.

\subsection{Benchmark Metrics}
\label{benchmarkmetrics}
We propose five benchmark metrics to enable comparison of feature toggle management practices across software projects, namely \textbf{toggle density} (prevalence), \textbf{churn rate} with \textbf{net accumulation} (growth dynamics), and \textbf{cleanup ratio} with \textbf{normalized lifespan} (lifecycle management).

For a project $p \in \mathcal{P}$, where $\mathcal{P} = \{\text{Kubernetes}, \text{GitLab}, \ldots\}$, let $|\mathcal{A}^{(p)}|$ denote the number of active feature toggles, $|\text{additions}^{(p)}|$ the total toggles added, $|\text{removals}^{(p)}|$ the total toggles removed, $L^{(p)}$ the lines of code, $T^{(p)}$ the analysis period in months, and $\tau^{(p)}$ the average release cycle duration (in days).

\paragraph{\textbf{C}hurn Rate.}
The first benchmark metric is the churn rate ($C^{(p)}$). It quantifies the total monthly toggle activity by summing additions and removals. This captures how actively a project manages its toggle inventory, with higher values indicating more frequent experimentation and cleanup cycles. It is computed as:

\begin{equation}
C^{(p)} = \frac{|\text{additions}^{(p)}| + |\text{removals}^{(p)}|}{T^{(p)}}
\end{equation}

\paragraph{\textbf{N}et Accumulation.}
We then propose using net accumulation ($N^{(p)}$), which measures the difference between addition and removal rates, indicating whether toggles are accumulating in the codebase over time. A positive value signals that toggle cleanup is not keeping pace with new toggle additions, potentially leading to technical debt. Conversely, a negative value indicates active debt reduction, where more toggles are removed than added. It is computed as:

\begin{equation}
N^{(p)} = \frac{|\text{additions}^{(p)}| - |\text{removals}^{(p)}|}{T^{(p)}}
\end{equation}

\paragraph{Cleanup \textbf{R}atio.}
The third metric is the cleanup ratio ($R^{(p)}$), which measures the proportion of added toggles that are eventually removed, indicating historical cleanup discipline in a software system. Since both $|\text{additions}^{(p)}|$ and $|\text{removals}^{(p)}|$ are cumulative counts over the full analysis period $T^{(p)}$, this ratio reflects overall cleanup effectiveness rather than a single-point snapshot. Equivalently, it can be expressed in terms of the previous two metrics, as: 

\begin{equation}
R^{(p)} = \frac{C^{(p)} - N^{(p)}}{C^{(p)} + N^{(p)}} = \frac{|\text{removals}^{(p)}|}{|\text{additions}^{(p)}|}
\end{equation}

\paragraph{Toggle \textbf{D}ensity.}
The fourth metric, toggle density ($D^{(p)}$), measures the number of active feature toggles per thousand lines of code (kLoC), where $|\mathcal{A}^{(p)}|$ and $L^{(p)}$ are both measured at a specific point in time, enabling the comparison across codebases of different sizes. We normalize by code size because it directly reflects the conditional code surface developers must navigate, while alternative normalizations such as project activity could capture complementary dimensions and merit future investigation. It is computed as:

\begin{equation}
D^{(p)} = \frac{|\mathcal{A}^{(p)}|}{L^{(p)}} \times 1000
\end{equation}

\paragraph{Toggle Life\textbf{S}pan.}
The last metric, normalized lifespan ($S_{\text{norm}}^{(p)}$), measures toggle duration in release cycles (rather than absolute days). This normalization reflects the fact that toggles are typically removed at release milestones, making release cycles a more meaningful unit for cross-project comparison than calendar days.

Let $\tilde{S}^{(p)} = \text{median}(\mathcal{S}^{(p)})$ denote the median lifespan (in days) of all removed toggles in project $p$, where $\mathcal{S}^{(p)} = \{S_1^{(p)}, S_2^{(p)}, \ldots, S_n^{(p)}\}$ is the set of individual toggle lifespans. Let $\tau^{(p)}$ denote the average release cycle duration (in days) for project $p$. The normalized lifespan is then:
\begin{equation}
S_{\text{norm}}^{(p)} = \frac{\tilde{S}^{(p)}}{\tau^{(p)}}
\end{equation}

By computing these five metrics, practitioners gain a quantitative basis for assessing feature toggle health. The following subsection establishes threshold zones for each metric, providing interpretive context for the computed values.

\begin{table}[t]
\centering
\small
\caption{Feature toggle interpretation framework, with proposed thresholds for toggle inventory health assessment. Its application to \textcolor{blue}{$\blacklozenge$}~Kubernetes and \textcolor{orange}{$\blacksquare$}~GitLab}
\label{tab:thresholds}
\begin{tabular}{@{}llllp{1.8cm}@{}}
\toprule
\textbf{Metric} & \textbf{Threshold} & \textbf{Zone} & \textbf{Description} & \textbf{Projects} \\
\midrule
\multirow{3}{*}{\shortstack[l]{Churn Rate\\($C^{(p)}$)}} 
    & $< 15$/month & Low & Deliberate changes & \textcolor{blue}{$\blacklozenge$}~\scriptsize{10.2} \\
    & $15$--$100$/month & Moderate & Balanced activity & \\
    & $\geq 100$/month & High & Rapid iteration & \textcolor{orange}{$\blacksquare$}~\scriptsize{104.5} \\
\midrule
\multirow{3}{*}{\shortstack[l]{Net Accumulation\\($N^{(p)}$)}} 
    & $< 2$/month & Sustainable & Cleanup keeps pace & \textcolor{blue}{$\blacklozenge$}~\scriptsize{1.5} \\
    & $2$--$5$/month & Warning & Gradual debt & \\
    & $\geq 5$/month & Critical & \textit{One-in-one-out} needed & \textcolor{orange}{$\blacksquare$}~\scriptsize{6.5} \\
\midrule
\multirow{3}{*}{\shortstack[l]{Cleanup Ratio\\($R^{(p)}$)}} 
    & $\geq 0.85$ & Healthy & $\geq$85\% removed & \textcolor{orange}{$\blacksquare$}~\scriptsize{0.88} \\
    & $0.70$--$0.85$ & Warning & Potential debt & \textcolor{blue}{$\blacklozenge$}~\scriptsize{0.74} \\
    & $< 0.70$ & Critical & Significant debt & \\
\midrule
\multirow{3}{*}{\shortstack[l]{Toggle Density\\($D^{(p)}$)}} 
    & $< 0.02$/kLoC & Conservative & Low toggle footprint & \textcolor{blue}{$\blacklozenge$}~\scriptsize{0.016} \\
    & $0.02$--$0.10$/kLoC & Moderate & Typical density & \textcolor{orange}{$\blacksquare$}~\scriptsize{0.081} \\
    & $\geq 0.10$/kLoC & Aggressive & Strict cleanup needed & \\
\midrule
\multirow{3}{*}{\shortstack[l]{Norm.\ LifeSpan\\($S_{\text{norm}}^{(p)}$)}} 
    & $< 3$ cycles & Short-lived & Rapid cleanup & \\
    & $3$--$8$ cycles & Moderate & Typical lifecycle & \textcolor{blue}{$\blacklozenge$}~\scriptsize{6.1}~\textcolor{orange}{$\blacksquare$}~\scriptsize{6.2} \\
    & $\geq 8$ cycles & Long-lived & Extended maintenance & \\
\bottomrule
\end{tabular}
\end{table}

\subsection{Interpretation Framework}
\label{empiricalthresholds}
While the five proposed metrics provide a quantitative view of a project's toggle management practices, the raw numbers alone do not say much without something to compare them to. Practitioners, for instance, cannot easily judge whether a given toggle density or churn rate in their system reflects healthy practices. To make these results more meaningful, we introduce a framework for interpreting feature toggle metrics.  We construct this framework by calculating all five metrics for Kubernetes and GitLab using the data from~\Cref{results}, and then defining threshold zones that treat these two systems as reference baselines.

To illustrate how we established these thresholds, consider the \textit{Net Accumulation} metric ($N^{(p)}$).  We regard a project as maintaining \textit{sustainable} toggle management if fewer than two toggles accumulate per month, a threshold that Kubernetes meets at 1.5 toggles per month. However, when the accumulation rate reaches five or more toggles per month, the project enters a \textit{critical} zone where toggles risk becoming technical debt, and a stricter cleanup policy, such as a ``one-in-one-out'' approach~\cite{hodgson2017feature}, becomes necessary. We applied similar reasoning to the other four metrics, basically using Kubernetes as a reference for \textit{conservative} toggle management and GitLab for more \textit{aggressive} practices.
\Cref{tab:thresholds} summarizes the resulting thresholds along with their interpretations. 

Using the framework is straightforward. Basically, to evaluate a project’s toggle health, practitioners can compute the five metrics from the version control history and compare the results with the thresholds in~\Cref{tab:thresholds}. If a project consistently falls within the conservative zone across all metrics, it likely follows a Kubernetes approach characterized by a low volume of feature toggles. In contrast, if it falls within the aggressive zone, it resembles GitLab’s high-churn toggle management approach. Mixed classifications (for example, high churn but a low cleanup ratio) may reveal process imbalances, such as toggle accumulation, that call for specific intervention, such as for instance cleanup.

These thresholds are intended as starting points for system self-assessment rather than strict rules. Since they are derived from two open source systems that may not represent the full spectrum of toggle management practices, teams should interpret them relative to their own project context. We expect that as more data become available from additional systems, the thresholds for the first four metrics can be refined to better reflect general industrial practices. In contrast, the last metric, normalized lifespan inherently adapts to each system's release cycle, making it self-calibrating by design.

\subsection{Applying the Framework}
The last column of~\Cref{tab:thresholds} shows where Kubernetes (\textcolor{blue}{$\blacklozenge$}) and GitLab (\textcolor{orange}{$\blacksquare$}) fall within each threshold zone, illustrating how the framework differentiates their toggle management approaches. Below, we show how the framework applies to each project and how this application can inform its use in other projects.

As it can be noticed, Kubernetes exhibits relatively modest toggle activity, with about ten toggle related events (additions and removals) per month ($C^{(p)} = 10.2$), placing it firmly within the \textit{low} churn zone. Its toggle inventory grows slowly, adding roughly one and a half more toggles than it removes each month ($N^{(p)} = 1.5$), which falls comfortably within the \textit{sustainable} range. Then, the codebase also maintains a sparse toggle footprint, with approximately one toggle per 64~kLoC ($D^{(p)} = 0.016$/kLoC), well within the \textit{conservative} zone. This overall pattern reflects a deliberate, release oriented strategy in which toggles are introduced sparingly but kept alive for extended periods.
 
GitLab, in contrast, operates at a much higher intensity. The project records over one hundred toggle events per month ($C^{(p)} = 104.5$), about ten times more than Kubernetes, placing it in the \textit{high} churn zone. Its toggle inventory grows rapidly, with a net accumulation of around six and a half toggles per month ($N^{(p)} = 6.5$), crossing into the \textit{critical} threshold. Despite this aggressive pace, GitLab demonstrates stronger cleanup discipline. Basically, 88\% of all introduced toggles are eventually removed ($R^{(p)} = 0.88$), corresponding to a \textit{healthy} status compared to Kubernetes's 74\%, which places it in the \textit{warning} zone. This apparent paradox, higher cleanup rate yet greater accumulation, can be explained by scale, namely GitLab adds nearly ten times as many toggles, so even with effective cleanup, the absolute number of remaining toggles remains higher. 

Perhaps most revealing is the normalized lifespan metric. In absolute terms, Kubernetes toggles persist nearly four times longer than those in GitLab (734 vs.\ 185 days). However, when expressed relative to each project's release cycle, both maintain toggles for approximately six release cycles ($S_{\text{norm}}^{(p)} \approx 6.1$ for Kubernetes with quarterly releases, $\approx 6.2$ for GitLab with monthly releases). This convergence suggests that despite vastly different behaviors, both projects share a common underlying practice, that is toggles typically span about six releases before removal, regardless of whether those releases occur monthly or quarterly.

These data reveal two distinct but equally valid approaches for feature toggle management. The first is a \textit{low-volume, long-tail} approach, exemplified by Kubernetes, where toggles are introduced sparingly, maintained across multiple release cycles, and removed through longer deprecation windows. 
The second is a \textit{high-volume, short-lived} approach, exemplified by GitLab, where toggles serve as lightweight, disposable switches that enable rapid feature iteration and A/B testing, but require aggressive cleanup discipline to prevent inventory accumulation. 
Our findings do not indicate that either approach is inherently superior. We believe that each reflects organizational practices and development rhythms tailored to the project's domain, release cycle, and acceptable failure cost.

\subsection{Interactive Benchmarking Dashboard}
\label{dashboard}
To support adoption of the proposed framework, we created an interactive web-based dashboard, which is available at \href{https://ternava.github.io/detog/}{https://ternava.github.io/detog/}.

The dashboard supports three main use cases. For \textit{visual comparison}, practitioners can inspect how Kubernetes and GitLab score across the five benchmark metrics on a radar chart, illustrating low-volume, long-tail versus high-volume, short-lived toggle management practices. For \textit{self-assessment}, teams can enter their own project metrics using the provided web form and see where they fall within the threshold zones, highlighting potential areas of concern. Finally, for \textit{community contribution}, users can submit and save new project data to a .csv file, enabling the benchmarking dataset to grow and the threshold boundaries to be refined in the future as more evidence is collected.

Providing such a framework through an interactive dashboard lowers the barrier for practitioners looking to assess and improve their toggle management practices. Then, as more projects are added, it can become an evolving collection of feature toggle lifecycle benchmarks
across diverse software ecosystems.

\section{Threats to Validity}
\label{threatstovalidity}
%

\paragraph{Internal Validity.}
\label{internalthreats}
One internal threat concerns \emph{toggle identification} accuracy. Our mining approach relies on pattern matching within specific files and directories. A small number of toggles defined in other Kubernetes files may be excluded from our analysis. However, manual inspection confirmed that these analysed locations represent the main toggle definition places in both projects and that any omissions are minimal.
Another threat involves \emph{refactoring events}. Bulk additions and removals (\emph{e.g.,} syntax changes, reordering, merging) can distort our findings. We identified three such events in Kubernetes and one in GitLab. While these spikes affect monthly counts, they do not impact much lifespan calculations since refactored toggles retain their original creation dates.
A further threat relates to \textit{negative lifespans}. A small number of toggles exhibit negative computed lifespans, likely due to git history anomalies such as rebasing, cherry-picking across branches, or timezone inconsistencies in commit metadata. Given their negligible count (3 in Kubernetes, 8 in GitLab), we retained these outliers, which do not significantly affect our lifespan analyses.

\paragraph{Construct Validity.}
\label{constructvalidity}
The five metrics in our benchmarking framework capture quantitative aspects of toggle lifecycle management. However, they may not fully represent toggle \textit{health}. Factors like toggle complexity, code coupling, test coverage of toggle paths, and documentation quality also affect maintainability but are not included. Future work could extend our framework to incorporate these qualitative dimensions.

\paragraph{External Validity.}
\label{externalthreats}
A first threat is related to \textit{generalizability}. Our study is based on only two open source systems with different architectures, 
sizes, and toggle management practices. Although Kubernetes and GitLab are large, and well-documented projects, the results may not generalize to closed source software, smaller codebases, or projects with different release cycles and development workflows.
Additionally, the increasing adoption of AI-assisted development tools may alter toggle dynamics and proposed thresholds in ways not captured by our historical data.
Then, \textit{toggle type homogeneity} is another threat. In both systems, feature toggles are only Boolean switches. Projects that rely on multi-valued toggles, percentage-based rollouts, or toggles with more explicit expiration metadata may show different lifecycle patterns that we did not cover in this study.
The \textit{time span} of our data may poses another threat. Our analysis covers 8.5 years of history for Kubernetes and 5 years for GitLab, and then toggle management practices can change over time. As a result, the observed patterns may reflect more mature processes rather than the intrinsic properties of toggles themselves. Therefore, replicating this study on additional systems and time periods would help strengthen the generalizability of our findings.

\section{Related Work}
\label{relatedwork}
\paragraph{Toggle Practices, Usage, and Complexity.}
Feature toggle research has expanded rapidly as teams increasingly rely on toggles for runtime variability management. Meinicke~\textit{et al.}~\cite{meinicke2020exploring} contrasted feature toggles with configuration options, clarifying when each mechanism is appropriate. 
On the practice side, Mahdavi-Hezaveh~\textit{et al.}~\cite{Hezaveh2021} identified 17~practitioner practices and proposed a toggle lifecycle model, while their subsequent work~\cite{mahdavihezaveh2022heuristics} contributed heuristics and code-level metrics for structuring toggles. Several empirical studies characterise the footprint of toggles in real systems. For instance, Meinicke~\textit{et al.}~\cite{meinicke2020capture} measured the prevalence of feature flags across open source projects, Rahman~\textit{et al.}~\cite{rahman2023feature} documented usage patterns in Chromium, and Prutchi~\textit{et al.}~\cite{prutchi2021adoption} showed that toggle adoption correlates with merge frequency and defect rates. At the code level, Rahman~\textit{et al.}~\cite{rahman2024exploring}, Schr{\"o}der~\textit{et al.}~\cite{schroder2022discovering}, and Tërnava~\textit{et al.}~\cite{ternava2022interaction} explored the complexity and interdependencies that toggle code introduces, while Ega and Motamarri~\cite{ega2025featureflags} discussed the tension between the flexibility toggles provide and the maintainability costs they incur. Collectively, this literature addresses toggle \textit{structure}, \textit{prevalence}, and \textit{complexity}, but not how toggle inventories evolve longitudinally or how lifespans compare across organizational contexts, which is the focus of our study.

\paragraph{Toggle Removal and Tooling.}
The risk that stale toggles accumulate as technical debt has been recognised in both academic and practitioner literature~\cite{daviestoggles20182,hodgson2017feature,neely2013continuous,unleash20212}. The most closely related empirical work is by Hoyos~\textit{et al.}~\cite{hoyos2021removal}, who studied toggle removal in 12~Python projects and surveyed 61~practitioners on their cleanup motivations. They showed that not all toggles are removed and that long-lived ones suggest technical debt. However, their analysis is cross-sectional rather than tracing accumulation and lifespan trajectories over multi-year time scales as we do here. On the tooling side, Piranha by Ramanathan~\textit{et al.}~\cite{ramanathan2020piranha} deletes stale flag code at Uber, Ketkar~\textit{et al.}~\cite{ketkar2024piranha} later extended it into a polyglot code-transformation language, Reflag~\cite{reflag2025aiflagcleanup} applies AI-assisted cleanup, and Shackleton~\textit{et al.}~\cite{shackleton2023dead} reported on large-scale automated dead-code removal at Meta. Our work complements these efforts by providing the empirical lifecycle data, toggle growth rates, accumulation dynamics, and survival distributions, needed to determine when and how aggressively such cleanup should be triggered.

\paragraph{Variability in Configurable Systems.}
Hunsen~\textit{et al.}~\cite{hunsen2016preprocessor} empirically characterised preprocessor-based variability in open-source and industrial systems, finding that \texttt{\#ifdef} usage grows as systems evolve. Kuiter~\textit{et al.}~\cite{sundermann2025linux} analysed two decades of the Linux kernel feature model history, tracking how the number of features has grown linearly at roughly 840 new features per year since 2002. Wolfart~\textit{et al.}~\cite{wolfart2021variabilitydebt} formalised such unchecked variability growth as \textit{variability debt}, a direct parallel to the toggle debt we quantify here. Jézéquel~\textit{et al.}~\cite{jezequel2022feature} then explicitly bridged compile-time and runtime variability, mapping SPL feature models directly to feature toggles. In contrast to these studies of feature count growth, our work focuses on the runtime dimension, measuring the toggle additions, removals, accumulation dynamics, and lifespan distributions across projects.

\section{Conclusion}
\label{conclusion}
We presented the first longitudinal study of feature toggle dynamics (prevalence, growth, and lifespan) in large-scale software systems. By analysing over 4,000 toggle events in Kubernetes (8.5 years) and GitLab (5 years), we provided empirical evidence on how many toggles are added, removed, and persist over time.

Our findings reveal four key insights. First, feature toggles are prevalent in modern codebases, with active inventories reaching hundreds of toggles (155 in Kubernetes, 403 in GitLab). Secondly, toggle removals consistently lag behind additions, leading to gradual inventory growth in software systems (35\% surplus in Kubernetes, 13\% in GitLab). Thirdly, toggle lifecycles are highly project-dependent, with median lifespans differing by nearly four times across software systems (734 days in Kubernetes vs. 185 in GitLab), reflecting fundamentally different management approaches. Lastly, a small but concerning subset of toggles tend to become permanent in the codebase (8 in Kubernetes, 25 in GitLab).

Building on these findings, we proposed a benchmarking framework with five metrics and empirically derived thresholds. To facilitate adoption, we also developed an interactive dashboard for self-assessment and community contributions.

Future work could strengthen these findings in several ways. Extending the dataset to additional systems would help validate and refine the proposed thresholds, including sensitivity analysis across different project sizes and domains. Investigating how AI-assisted development affects toggle dynamics is also increasingly relevant as such tools become widespread. On the practical side, building CI/CD tools that flag toggles approaching critical thresholds and automating their removal once they expire could help teams manage toggle debt proactively. Finally, incorporating qualitative factors like toggle complexity and documentation quality, complemented by practitioner interviews to validate whether the proposed thresholds align with real-world cleanup decisions and to uncover organizational root causes behind toggle accumulation, would provide a more comprehensive view of toggle health.

Ultimately, we hope this work helps teams make informed decisions about their toggle practices and encourages broader efforts to establish common standards for when to add, how long to keep, and when to remove feature toggles.

%
%
\bibliographystyle{splncs04}
\bibliography{bibliography}

\end{document}